\documentclass[twocolumn,showpacs,prl,footinbib,a4paper,superscriptaddress,floatfix]{revtex4}

\usepackage{amssymb,amsmath}
\usepackage{graphics}
\usepackage{dcolumn}
\usepackage{bm}
\usepackage{epsfig}

\newcommand{\expect}[1]{\langle #1 \rangle}

\usepackage[usenames,dvipsnames]{color}
\definecolor{MyOrange}{rgb}{1.0,0.5,0}
\definecolor{MyPurple}{rgb}{0.5,0,1}

\begin{document}

\title{Nonequilibrium Fluctuation-Dissipation Theorems for Interacting Quantum Transport}

\author{H. Ness}
\email{herve.ness@york.ac.uk}
\altaffiliation{European Theoretical Spectroscopy Facility (ETSF)}
\affiliation{Department of Physics, University of York, Heslington, York YO10 5DD,UK}

\author{L. K. Dash}
\altaffiliation{European Theoretical Spectroscopy Facility (ETSF)}
\affiliation{Department of Physics and Astronomy, University College London, Gower Street, London WC1E 6BT, UK}

\begin{abstract}
  We study non-equilibrium (NE) fluctuation-dissipation (FD) relations in the context of 
  quantum thermoelectric transport through a two-terminal nanodevice, in the steady-state and with interaction.
  The FD relations for the one- and two-particle correlation functions are derived. 
  Numerical applications, using self-consistent NE Green's functions calculations, are given for electron-phonon 
  interaction  in  the central region.
  We find that the FD relations for the one-particle correlation function are strongly dependent on 
  both the NE conditions and the interactions, while they are much less dependent on the interactions 
  for the two-particle correlation. This suggests interesting applications
  for single-molecule and other nanoscale transport experiments:
  the two-particle correlation functions, obtained from noise and transport measurement,
  provide information about the gradients of chemical potential and
  temperature, and other properties of the system.
 
\end{abstract}
 

\maketitle

Quantum systems can be driven far from equilibrium by time-dependent perturbation
or by coupling to reservoirs at different chemical potentials or temperatures.
In the latter case, the system is ``open'' and particle- or energy-currents flow 
throughout the system. 
Such processes take place in different contexts, ranging from nanoscale quantum 
transport to chemical reactions.
The recent developments in modern techniques of microscopic manipulation and nanotechnologies
enable us to build functional nanoscale systems.
Fluctuations in such systems can nowadays be measured at the single-electron 
level \cite{Blanter:2000,Forster:2008}.
At equilibrium, small fluctuations satisfy a universal relation
known as the fluctuation-dissipation (FD) theorem \cite{Kubo:1957,Kubo:1966,Efremov:1969}.
The FD relation connects spontaneous fluctuations to the linear response, 
for both classical and quantum systems. 

The search for similar relations for systems driven far from equilibrium has been an 
active area of research for many decades. A major advance had taken place with 
the derivation of exact fluctuation relations which hold for classical and quantum systems 
at non-equilibrium (NE) \cite{Bochkov:1977,Bochkov:1979,Jarzynski:1997,Jarzynski:2004}.
For quantum systems, fluctuations have also been studied in the context of quantum heat conduction 
and full counting statistics 
\cite{Saito:2007,Andrieux:2008,Gelin:2008,Andrieux:2009,Esposito:2009,Talkner:2009,Flindt:2010,Campisi:2011,Safi:2011}.
Another route is to consider the equilibrium relations with
effective and local thermodynamical variables (temperature, chemical potential) dependent on 
the NE conditions \cite{Arrachea:2005,Caso:2012}.

In this paper, we focus on the generalisation of FD relations to NE conditions in the presence 
of both charge and heat transport, with a strong emphasis on the effects of interactions between 
particles on such relations.
In particular, we derive the NE FD relations for 1-particle correlation functions  
(the electronic Green's functions GFs) and for 2-particle correlation functions 
(the charge-charge (CC) and current-current (JJ) correlation and response functions).
We calculate the Kubo-Martin-Schwinger (KMS) and FD relations for a model system connected 
to two reservoirs in the presence of an applied bias and a temperature gradient,
in the NE steady-state. 

We show that the FD theorem for the 1-particle correlation functions is strongly dependent on
both the NE conditions and the interaction between particles. While the FD relations for the
2-particle correlation functions are much less dependent on the interaction. 
Such 2-particle quantities are accessible experimentally by noise and transport measurements. 
Hence one could determine properties of the system such as the effective temperature, 
the gradients of temperature and chemical potential, the strength of the coupling
to the leads, and deviations from electron-hole symmetry.
 
\emph{Equilibrium FD theorems.---}
At equilibrium, the FD theorem arises from the fact that the time evolution operator
$e^{-iHt}$ bears a strong formal similarity to the weighting factor $e^{-\beta H}$
that occurs in the statistical averages by identifying $t \equiv -i \beta$ (with $\beta=1/kT$). 
The key relation is that, for any two operators $A$ and $B$, one has
$\langle A(t-i\beta) B(t') \rangle = \langle B(t') A(t) \rangle$. 

We can define the quantity $X^>_{AB} = \langle A(1) B(2) \rangle$ and
$X^<_{AB} = \mp \langle B(2) A(1) \rangle$, with the minus (plus) sign for $A,B$ being 
fermion (boson) operators. At equilibrium, the quantities depend only on the time difference,
and after Fourier transform (FT), we write the general FD theorem for $X^\lessgtr_{AB}(\omega)$
as
\begin{equation}
\label{eq:equiFDT}
\begin{split}
X^>_{AB}+X^<_{AB} 
= 
\left[ \frac{r_{AB}(\omega)+1}{r_{AB}(\omega)-1} \right] 
\left( X^>_{AB} - X^<_{AB} \right) \ ,
\end{split}
\end{equation}
with the ratio $r_{AB}(\omega)$ obtained from the KMS relation
at equilibrium $r_{AB}(\omega)=X^>_{AB}/X^<_{AB}= \mp e^{\beta\bar\omega}$ with the
minus sign for fermion operators and $\bar\omega=\omega$ ($\bar\omega=\omega-\mu^{\rm eq}$) for 
(grand-)canonical average (with the equilibrium Fermi level $\mu^{\rm eq}$),
and with the plus sign for boson operators ($\bar\omega=\omega)$.
One recovers the conventional FD relations from Eq.(\ref{eq:equiFDT}). For boson 
operators, the usual relation between commutator and anticommutator is
\begin{equation}
\label{eq:equiFDT_boson}
\begin{split}
\langle \left\{ A, B \right\} \rangle_\omega
= 
\coth \left( {\beta\bar\omega}/{2} \right) 
\langle \left[ A, B \right] \rangle_\omega \ ,
\end{split}
\end{equation}
with $\langle \dots \rangle_\omega$ being the FT
of $\langle \dots \rangle(t-t')$.
For electron Green's functions, $i G^>(1,2)= X^>_{AB}$ with $A(1)=\Psi(1)$
and $B(2)=\Psi^\dag(2)$, and we recover the usual relation
$G^K(\omega)= \tanh \left( \beta\bar\omega/2 \right) 
\left( G^r - G^a \right)(\omega)  $
by identifying $G^K=G^>+G^<=i\langle [\Psi(1) , \Psi^\dag(2)] \rangle$ and
$G^r-G^a=G^>-G^<=i\langle \{\Psi(1) , \Psi^\dag(2)\} \rangle$ \cite{Note:0}.

\emph{NE steady-state transport.---}
The above KMS and FD relations however do not hold in NE 
conditions, even in the steady state which can be seen as a pseudo-equilibrium 
state \cite{Hershfield:1993,Doyon:2006,Han:2007,Fujii:2007,Han:2010}.
We now extent the FD relations to NE steady-state quantum transport. 
We consider the single impurity model connected to two non-interacting Fermi seas. 
The left ($L$) and right ($R$) leads are at their own equilibrium, with a Fermi 
distribution $f_\alpha(\omega)$ defined by their respective 
chemical potentials $\mu_\alpha$ and temperatures $T_\alpha$ ($\alpha=L,R$).
The central region connected to the leads contain interaction
characterized by a self-energy  $\Sigma_{\rm int}(\omega)$ \cite{Note:1}.
Furthermore the specific model used for the leads does not need
to be specified, as long as the leads can be described by an embedding
self-energy $\Sigma_{\alpha}(\omega)$ in the electron GF of the central region.
Our results for the FD theorem are general with respect to both the leads and
the interaction self-energies.

\emph{FD relations for the 1-particle correlation functions.---}
In the absence of interaction in the central region, we use the
properties of the GFs in the central region, 
$G_0^\lessgtr(\omega)=G_0^r(\omega) (\Sigma^\lessgtr_{L}+\Sigma^\lessgtr_{R})(\omega) G_0^a(\omega)$,
to show that they follow a pseudo-equilibrium relation \cite{Hershfield:1991}:
$G_0^\lessgtr(\omega)  = -  f^\lessgtr(\omega)    \left( G_0^r - G_0^a \right)(\omega)$,
where the NE distribution $f^<=f_0^{\rm NE}$ and $f^>=f_0^{\rm NE}-1$ 
with $f_0^{\rm NE}(\omega)=(\Gamma_L(\omega)f_L(\omega)+\Gamma_R(\omega) f_R(\omega))/\Gamma$
and $\Gamma=\Gamma_L+\Gamma_R$ with $\Gamma_\alpha(\omega)=i(\Sigma_\alpha^>-\Sigma_\alpha^<)(\omega)$.
We determine the FD ratio (FDr) from Eq.~(\ref{eq:equiFDT}) as
${\rm FDr}[G_0]=(G_0^>+G_0^<)/(G_0^>-G_0^<)=1-2f_0^{\rm NE}(\omega)$.
 
For symmetric coupling to the leads, $\Gamma_L=\Gamma_R$ and $\mu_{L,R}=\mu^{\rm eq}\pm V/2$, we find
the following expression \cite{Kirchner:2009}:
\begin{equation}
\label{eq:FDratioG0sym}
{\rm FDr}[G_0]=\frac{\sinh \beta\bar\omega}{\cosh \beta\bar\omega + \cosh \beta V/2} \ ,
\end{equation}
and for the KMS ratio $r_{AB}$:
\begin{equation}
\label{eq:KMSratioG0sym}
\frac{G_0^<}{G_0^>}=-\frac{e^{-\beta\bar\omega}+\cosh \beta V/2}{e^{\beta\bar\omega} + \cosh \beta V/2} \ .
\end{equation}
At equilibrium ($V=0$), we recover the usual results ${G_0^<}/{G_0^>}=-e^{-\beta\bar\omega}$
and ${\rm FDr}[G_0]=\tanh \left( \beta\bar\omega/2 \right)$.

For asymmetric contacts and potential drops $\mu_\alpha=\mu^{\rm eq}+\eta_\alpha V$
(with $\eta_L-\eta_R=1$) we find \cite{Note:2b}:
\begin{equation}
\label{eq:FDratioG0}
{\rm FDr}=\frac{\sinh \beta(\bar\omega-\bar\eta V) - (\bar\Gamma_L-\bar\Gamma_R) \sinh \beta V/2}
                        {\cosh \beta(\bar\omega-\bar\eta V) + \cosh \beta V/2} \ ,
\end{equation}
with $\bar\eta=(\eta_L+\eta_R)/2$ and $\bar\Gamma_\alpha=\Gamma_\alpha/\Gamma$. 

In the presence of interaction in the central region, with a self-energy 
$\Sigma_{\rm int}(\omega)$,
we use again the properties of the NE GF to
find that
\begin{equation}
\label{eq:KMSratioG}
\begin{split}
\frac{G^<}{G^>} 
                  =\frac{\Sigma^<_{L+R}+\Sigma^<_{\rm int}}{\Sigma^>_{L+R}+\Sigma^>_{\rm int}}
=\frac{f_0^{\rm NE} - i \Sigma^<_{\rm int}/\Gamma}{f_0^{\rm NE}-1 - i \Sigma^>_{\rm int}/\Gamma} \ .
\end{split}
\end{equation}
From this ratio, we define a NE distribution function $f^{\rm NE}(\omega)=[1-G^>/G^<]^{-1}$
which permits us to define the pseudo-equilibrium relation
$G^<  = -  f^{\rm NE}    \left( G^r - G^a \right)$.
It is given by
\begin{equation}
\label{eq:fNE}
f^{\rm NE}(\omega) = \frac{f_0^{\rm NE}(\omega) - i \Sigma^<_{\rm int}(\omega)/\Gamma(\omega)}
{1+i (\Sigma^>_{\rm int}-\Sigma^<_{\rm int} ) / \Gamma } \ .
\end{equation}
There is no \emph{a priori} reason for $f^{\rm NE}$ to be equal to the non-interacting distribution 
$f_0^{\rm NE}$ \cite{Note:2}.
From Eq.~(\ref{eq:fNE}), we derive the FDr $(1-2f^{\rm NE})$ for the interacting GFs
\begin{equation}
\label{eq:FDratioG}
{\rm FDr}[G]=\frac{ {\rm FDr}[G_0] + i ( \Sigma^>_{\rm int} + \Sigma^<_{\rm int} ) / \Gamma }
                      { 1 + i ( \Sigma^>_{\rm int} - \Sigma^<_{\rm int} ) / \Gamma          } \ .
\end{equation}

The NE FD relations, Eqs.(\ref{eq:FDratioG0}-\ref{eq:FDratioG}), derived for the
one-particle correlations
are less universal than 
the equilibrium relation, since they depend on both the set-up that drives the system
out of equilibrium and on the MB interaction, as expected. 
However the NE FD relations are universal, with respect to the interaction, in the same sense 
that the GFs have an universal expression via the use of the interaction self-energies.

We now provide a numerical application for a specific choice of interaction.
We consider a model system with electron-phonon (e-ph) interaction for which we calculate the full NE
properties using the NE GF Keldysh formalism \cite{Dash:2010,Dash:2011}.
The Hamiltonian for the central region is
$  H_C 
  = \varepsilon_0 d^\dagger d + \omega_0 a^\dagger a +
  \gamma_0 (a^\dagger + a) d^\dagger d $ ,
where $d^\dagger$ ($d$) creates (annihilates) an
electron in the level $\varepsilon_0$, which
is coupled to  the vibration mode 
of energy $\omega_0$ via the coupling constant $\gamma_0$.
The leads are represented by one-dimensional
tight-binding chains and $t_{0L,R}$ are the hopping integrals
to the central region \cite{Note:onleadSE,Note:onSupplMat}. 
The many-body (MB) e-ph interaction self-energies $\Sigma_{\rm int}$ are treated at the
self-consistently Born approximation level \cite{Dash:2010,Dash:2011,Note:onSupplMat}.

We have performed calculations for a wide range of parameters found in \cite{Note:onSupplMat}. 
We consider symmetric ($t_{0L}=t_{0R}$) and asymmetric ($t_{0L}\ne t_{0R}$)
coupling to the leads, different strengths of coupling $t_{0\alpha}$, transport regimes 
(off-resonant $\varepsilon_0 \gg \mu^{\rm eq}$, and
resonant  $\varepsilon_0 \sim \mu^{\rm eq}$), e-ph coupling strengths,
biases $V$ with symmetric and asymmetric potential drops at the contacts,
and temperatures $T_\alpha$ and $T_{\rm ph}$.  
All calculations \cite{Note:onSupplMat} corroborate the conclusions we find for the behaviour 
of the NE FD ratios that we present below for specific sets of parameters.

\begin{figure}
  \center
  \includegraphics[clip=,width=\columnwidth]{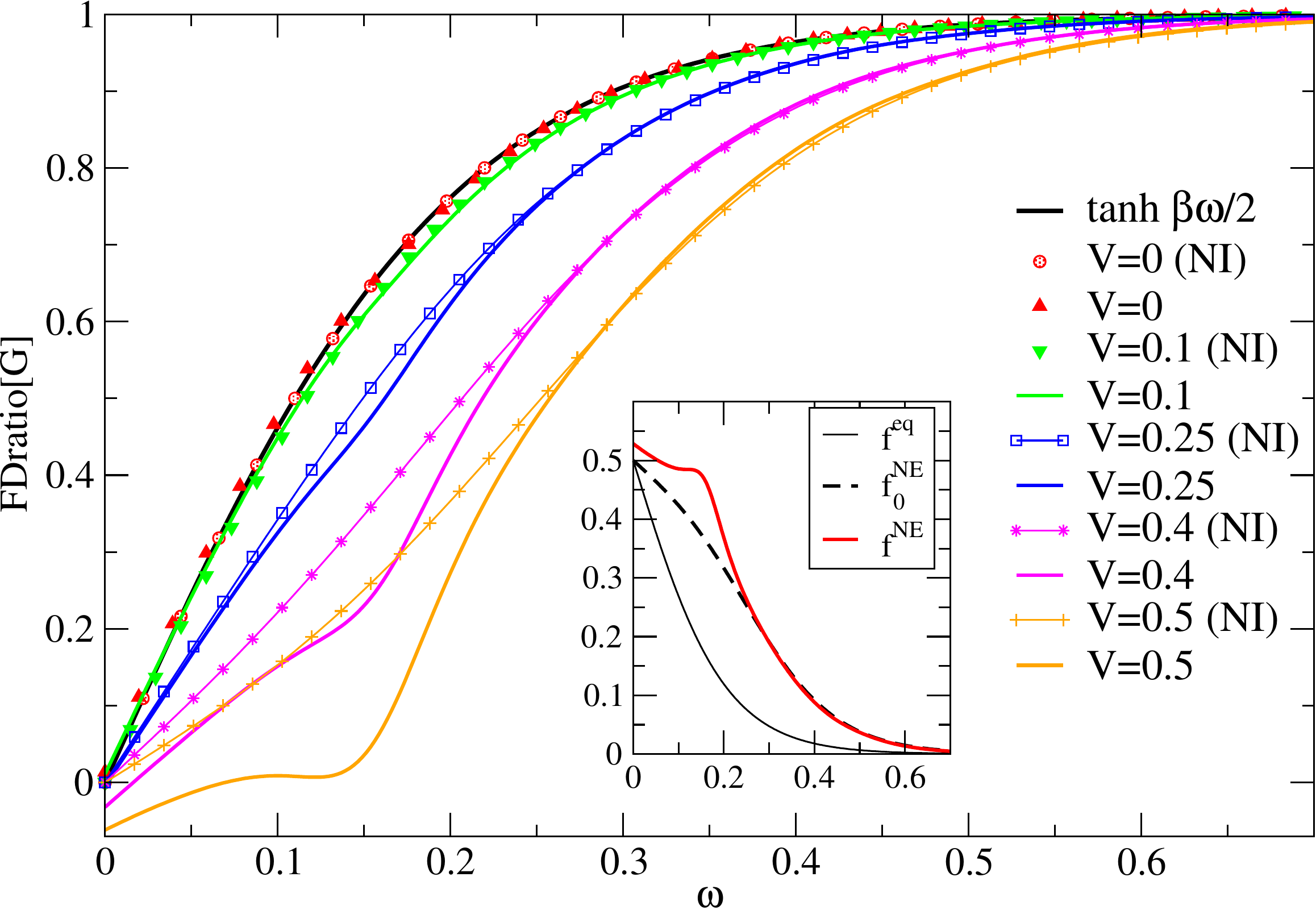}
  \caption{(color online)
FD ratio for the GF and both the non-interacting (NI) and interacting cases in the off-resonant
regime ($\varepsilon_0=0.50$) and for different biases $V$. At equilibrium, the
FD ratio is given by $\tanh \beta\omega/2$ (see $V=0$, and $V=0.1$).
The presence of interaction induces strong deviation from the non-interacting
FD ratio giving in Eq.~(\ref{eq:FDratioG0sym}).
The other parameters are $\gamma_0=0.12, \omega_0=0.3$,  
$t_{0\alpha}=0.15$, $T_\alpha=T_{\rm ph}=0.1$, $\eta_L=0.5$, 
$\varepsilon_\alpha=0, \beta_\alpha=2$.
Inset: Equilibrium distribution $f^{\rm eq}(\omega)$ and NE distributions 
$f_0^{\rm NE}$ and $f^{\rm NE}$ in the absence and presence of interaction.} 
  \label{fig:1}
\end{figure}

Figure \ref{fig:1} shows the FD ratio for the GF for the off-resonant transport regime, in both
the presence and the absence of interaction. For zero and very low bias, the FDr follows
the equilibrium $\tanh \beta\omega/2$ expression as expected. For the non-interacting case, the FD
ratio follows Eq.~(\ref{eq:FDratioG0sym}). One can clearly see that the presence of
interaction strongly modifies the FDr. The effects are stronger for larger $V$ when
the bias window include a substantial spectral weight of the self-energy $\Sigma_{\rm int}^\lessgtr$.
This is the regime when the single-(quasi)particle representation for quantum transport
breaks down \cite{Ness:2010}.
At large bias, we can obtain negative values of the FDr. This is when the NE MB effects
are not negligible and induce strong modifications of the NE distribution $f^{\rm NE}$. 
In that case $f^{\rm NE}>0.5$ for $\omega \ge 0$ (see inset in Fig.~\ref{fig:1}).
Such a behaviour never occurs in the resonant transport regime (with symmetric coupling to the leads 
and without Hartree-like self-energy) when the system always presents electron-hole 
symmetry \cite{Note:onSupplMat}.
These results show that the NE FD theorem for the GF is strongly dependent on the NE conditions 
as well as on the MB effects.
However the NE GF are not directly accessible experimentally as are the electronic current and charge
for which the dependence on both the NE and MB effects has been shown in \cite{Dash:2010,Dash:2011}.

\emph{FD relations for the 2-particle correlation functions.---}
We now calculate the FD relations for the JJ and CC correlation and response functions, far from equilibrium, 
and compare such relations with those obtained for the GFs.
By definition \cite{Blanter:2000}, the fluctuation correlation function (noise) is
$S^X_{\alpha\beta}(t,t')=\frac{1}{2}
\langle\left\{ \delta X_\alpha(t),\delta X_{\beta}(t') \right\}\rangle $
where $\delta X_\alpha(t)= X_\alpha(t)- \expect{X_\alpha}$.
The response function is
$R^X_{\alpha\beta}(t,t')=\langle\left[ X_\alpha(t),X_{\beta}(t') \right]\rangle$.
For the current flowing at the $\alpha=L,R$ contact: $X_\alpha=J_\alpha$ and
$\langle J_\alpha(t)\rangle = {e}/{\hbar}\ {\rm Tr}_\alpha \left[ (\Sigma G)^<(t,t) - (G \Sigma)^<(t,t) \right]$.
For the charge in the central region: $X_C=n_C$ and $\langle n_C(t)\rangle = 
{e}\ {\rm Tr} [ -i G^<(t,t) ]$.
The total noise $S^J(\omega)$ and response function $R^J(\omega)$ are defined from the symmetrized
current $J=(J_L-J_R)/2$.

In the steady state, all quantities depend only on the difference $t-t'$ and, after FT, we obtain
the relation
$\begin{array}{c} 2S^x(\omega) \\ R^x(\omega) \end{array} = \langle x x \rangle _\omega^- \pm \langle x x \rangle _\omega^+$ 
with
$\langle x x \rangle_\omega^\pm=\langle j j \rangle_\omega^\pm$
for the current, or $\langle n_C n_C \rangle _\omega^\pm$ for the charge \cite{Note:2.5,Note:3}.
The 2-particle correlation functions are bosonic. At equilibrium they follow the relation 
$2S^x(\omega) = \coth \left( {\beta\omega}/{2} \right) R^x(\omega)$ \cite{Note:4}.

We now compute the two-particle correlation functions and the FD relations using our model NEGF 
calculations with e-ph interaction.
We concentrate below on the inverse of the NE FD ratio $1/{\rm FDr}=R^x(\omega)/2S^x(\omega)$ 
for JJ and CC correlation functions. 
This allows us to avoid the divergence of the $\coth$-like function 
at $\omega=0$, and allows for a direct comparison with the FDr of the GFs 
(function $\tanh {\beta\omega}/{2}$ at equilibrium).
\begin{figure}
  \center
  \includegraphics[clip=,width=\columnwidth]{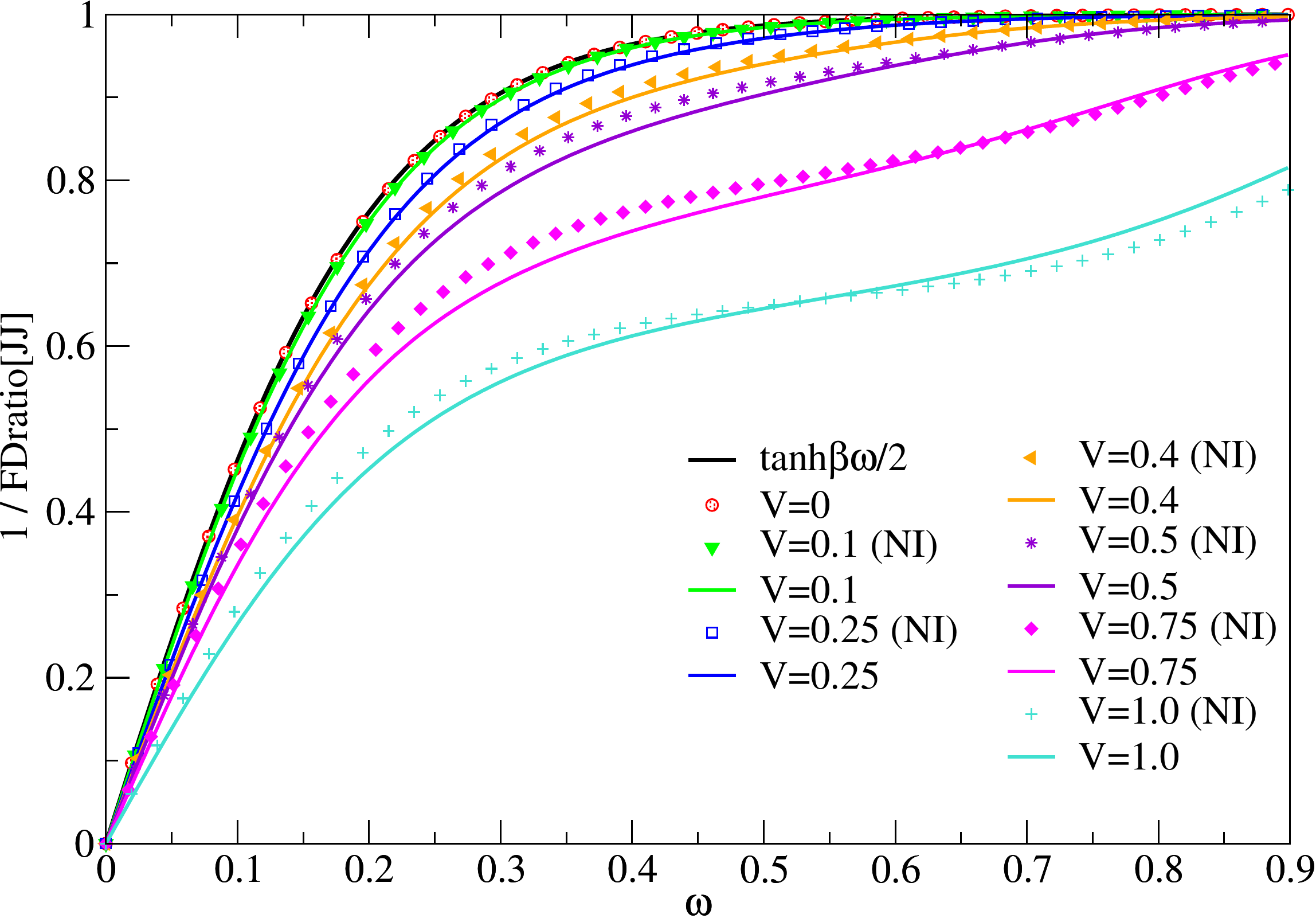}
  \caption{(color online)
Inverse FD ratio for current-current correlation function in the off-resonant regime, and
for the non-interacting (NI) and interacting cases.
The parameters are $\varepsilon_0=0.50$, $\gamma_0=0.12, \omega_0=0.3$,  
$t_{0\alpha}=0.15$, $T_\alpha=T_{\rm ph}=0.1$, $\eta_L=0.5$, 
$\varepsilon_\alpha=0, \beta_\alpha=2$.}
  \label{fig:3}
\end{figure}
Figure \ref{fig:3} shows the inverse FDr of the JJ correlation functions for
the off-resonant regime (same parameters as in Fig.~\ref{fig:1}) (for the resonant regime
see \cite{Note:onSupplMat}).
The inverse of FDr for the JJ correlation functions does not have the same
behaviour as the FDr of the GFs \cite{Note:onSupplMat}, although it follows the same
 $\tanh \beta\omega/2$ behaviour at very small applied bias, as expected.
Increasing the bias, seems to correspond to an effective increase of the temperature.
However the FDr[JJ] is never well represented by an 
$\tanh \beta_{\rm eff}\omega/2$ function (with an effective local temperature 
$kT_{\rm eff}=1/ \beta_{\rm eff}$) beyond the linear regime \cite{Note:onSupplMat}. 
More importantly, the NE FD ratio for the JJ correlations is much less dependent 
on the interaction than FDr[G]. This is a very interesting property, useful for experiments
as we explain below.

\begin{figure}
  \center
  \includegraphics[clip=,width=\columnwidth]{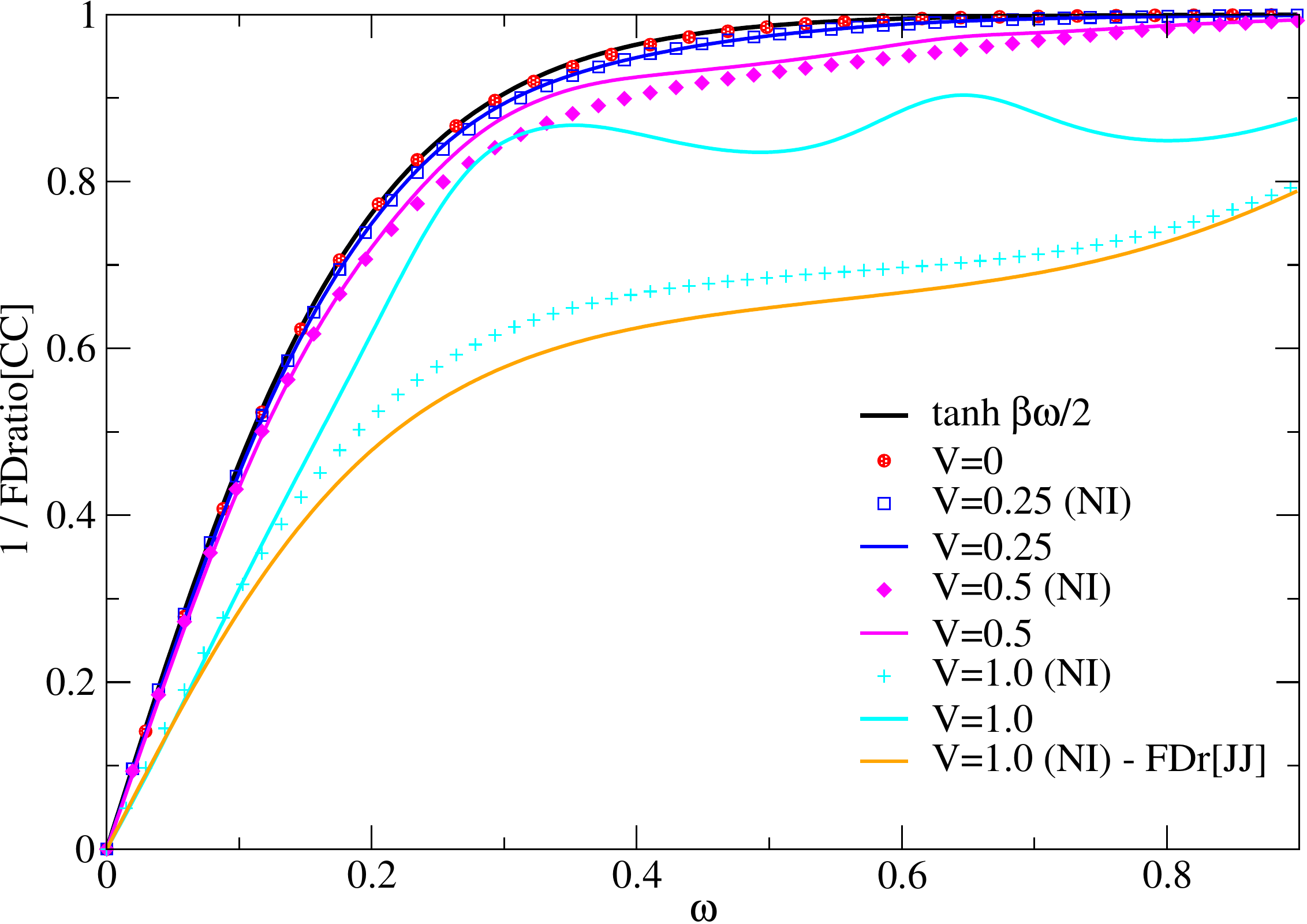}
  \caption{(color online)
Inverse FD ratio for CC correlation function in the off-resonant regime, and
for the non-interacting (NI) and interacting cases.
One curve for the FDr[JJ] is also given to show that FDr[CC] $\ne$ FDr[JJ].
The parameters are $\varepsilon_0=0.50$, $\gamma_0=0.12, \omega_0=0.3$,  
$t_{0\alpha}=0.15$, $T_\alpha=T_{\rm ph}=0.1$, $\eta_L=0.5$, 
$\varepsilon_\alpha=0, \beta_\alpha=2$.}
  \label{fig:4}
\end{figure}

Figure \ref{fig:4} shows the inverse FDr of the CC correlation functions in
the off-resonant regime (same parameters as in Fig.~\ref{fig:1}).
We observe again that the interaction effects are less dominant in FDr[CC] than in FDr[G], 
except for large bias.
Furthermore we find that FDr[CC] $\ne$ FDr[JJ] \cite{Note:onSupplMat}.
The reasons why the FD ratios for the JJ and the CC correlations and GFs are all different 
can be understood from the NE density matrix (including both NE and MB effects)  introduced by 
Hershfield as 
$\rho \equiv e^{-\beta (H-Y)}$ \cite{Hershfield:1993}.
The $Y$ operator is constructed from an iterative scheme for the equation
of motion of an initial $Y_0$ operator.
In the case of a two-terminal device, the initial operator is $Y_0=\mu_L N_L + \mu_R N_L$
with the left and right chemical potentials $\mu_{L,R}$ and particle number operators $N_{L,R}$.
The key relation leading the FD theorems becomes
$\langle A(t-i\beta) B(t') \rangle = \langle e^{-\beta Y } B(t') e^{\beta Y} A(t) \rangle$.
The usual equilibrium FD relations break down at NE and additional
contributions arise from the expansion
$ e^{-\beta Y } B e^{\beta Y} = B + [-\beta Y, B] + \dots$ \cite{Hershfield:1993,Tasaki:2006}.
For the GFs of the central region, the fermion operator $B$ 
is $d$ or $d^\dag$, while for the JJ (CC) correlations, the boson-like operator $B$ is
$c^\dag_\alpha d$ or $d^\dag c_\alpha$ ($d^\dag d$) where $c^\dag_\alpha$ ($c_\alpha$) creates
(annihilates) an electron in the lead $\alpha$. Therefore there is no obvious reason
for the two FD ratios to be identical, especially in the presence of interaction.
Furthermore, with the JJ and CC correlations, one deals with an higher order product than for the
GFs and the expansion of $ e^{-\beta Y } B e^{-\beta Y} $ contains higher order
powers in terms of the interaction coupling parameters (in case our case, in terms of $\gamma_0^2$)
in comparison to the series expansion for the GFs.
Therefore, for weak to intermediate interaction strengths, we expect less effect from 
the interaction in the FDr of the JJ and CC correlations than in FDr[G] as shown above.

\emph{With temperature gradient.---}
So far we have considered systems with a unique temperature. We also consider cases where
there is a temperature gradient between the $L$ and $R$ lead.
The derivations follow the same line as previously, and we find the NE FD ratio for the 
symmetric non-interacting case (in the presence of both a potential and temperature gradients between 
the two leads):
\begin{equation}
\label{eq:FDratioG0_withgradT}
\begin{split}
  {\rm FDr}[G_0] = & \left( 2\sinh \bar\beta\bar\omega \right) /
   \left( 2\cosh \bar\beta\bar\omega  
 + \left( u^{-\beta_L} + u^{\beta_R}\right) \times \right. \\
&  \left.  \cosh {\Delta\beta\bar\omega}/{2}  
 + \left( u^{-\beta_L} - u^{\beta_R}\right) \sinh {\Delta\beta\bar\omega}/{2}  \right)  , \nonumber
\end{split}
\end{equation}
with $u=e^{V/2}$, $\bar\beta=(\beta_L+\beta_R)/2$ and $\Delta\beta=\beta_L-\beta_R$.
Clearly, at zero bias, the gradient $\Delta\beta$ plays the same role as the bias
$V$ in Eq.~(\ref{eq:FDratioG0sym}) with an effective temperature $T_{\rm eff}$ defined from 
$\bar\beta$ as $T_{\rm eff}=T_L T_R / (T_L+T_R)$.
The calculations we have performed in the presence of both temperature gradient and applied bias show
similar behaviours as described above \cite{Note:onSupplMat}. Now we have two 
``forces'' driving the system out of equilibrium, $\Delta\mu$ and $\Delta\beta$, and the FD relations
(at low bias) are governed by the effective temperature $1/\bar\beta$.

\emph{Discussion.---}
We have derived FD relations for one-particle and two-particle correlation functions in
the context of quantum transport through a two-terminal device in the steady state regime.
We have also provided numerical application of our derivation for the case of a single
impurity model in the presence of e-ph interaction.
Our calculations are mostly relevant for e-ph interacting systems, but are 
not limited only to these processes. 
The main conclusion of our work is that the FD relations for the GFs are strongly dependent
both on the `forces' ($\Delta\mu$ and/or $\Delta\beta$) driving the system out of equilibrium
and on the interaction, while the FD relations for the current-current correlation functions
are much less dependent on the interaction.
However the JJ FD relations cannot be well described by equilibrium relations using an effective
local temperature (which would be dependent on the applied `forces').

The weak dependence on interaction of the JJ FD relation implies that the calculated
relation for the non-interacting case can serve as a master curve for fitting 
experimental results. The experimental JJ correlation functions obtained via noise and transport
measurement, fitted on the master curve, can provide us with information about the `forces' 
($\Delta\mu$ and/or $\Delta\beta$) and effective temperature in the central region. 
These are
crucial quantities to know in single-molecule nanodevice experiments.
Furthermore, a strong departure from the master curve could indicate a breakdown of the
major hypothesis of in our model, i.e. the interactions are not located only in the central
region, or there are more than two energy/particle reservoirs connected to the central region.

We thank L. Arrachea and A.J. Fisher for useful comments, and AJF for suggesting the calculation of the CC correlations.


\begin{thebibliography}{29}
\bibitem{Note:2b}
For the KMS ratio we obtain \\
${G_0^<}/{G_0^>}=-
\frac{e^{-\beta(\bar\omega-\bar\eta V)} + \cosh \beta V/2 + (\bar\Gamma_L-\bar\Gamma_R) \sinh \beta V/2}
     {e^{ \beta(\bar\omega-\bar\eta V)} + \cosh \beta V/2 - (\bar\Gamma_L-\bar\Gamma_R) \sinh \beta V/2} $.
\bibitem{Note:onleadSE}
The corresponding lead self-energies $\Sigma^x_\alpha(\omega)$ are energy dependent
and go beyond the wideband limit.
\bibitem{Note:2.5}
For the CC correlations we have: \\
$\langle n_C n_C \rangle _\omega^{\mp}= {e^2} \int {{\rm d}u}/{2\pi}\  {\rm Tr} [G^>(u) G^<(u\mp\omega ] $
with the trace taken on the electron states of the central region.
\end{thebibliography}
\end{document}